\pdfoutput=1
\documentclass[twoside,english]{mhd}
\usepackage[T1]{fontenc}
\usepackage{mathtools}
\usepackage{amsbsy}
\usepackage{amssymb}
\usepackage{cancel}
\usepackage{mathdots}
\usepackage{stmaryrd}
\usepackage{stackrel}
\PassOptionsToPackage{version=3}{mhchem}
\usepackage{mhchem}
\usepackage{esint}
\usepackage{babel}
\mhdhead{40}{1}{1}	
\newcommand{\apj}{ApJ}

\newcommand{\aap}{A~\&~A}
\newcommand{\mnras}{MNRAS}

\newcommand{\physrep}{Phys. Rep.}

\title{Magnetic helicity in non-axisymmetric mean-field solar dynamo.}

\author{V.V. Pipin}
\institute{Institute of Solar-Terrestrial Physics, Russian Academy of
Sciences}

\begin{document}

\maketitle

\begin{abstract}
The paper address the effects of magnetic helicity conservation in
a non-linear non-axisymmetric mean-field solar dynamo model. We study
the evolution of the shallow non-axisymmetric magnetic
field perturbation with the strength about 10G in the solar convection zone. The dynamo evolves from the
pure axisymmetric stage through the
short (about 2 years) transient phase when the non-axisymmetric m=1
dynamo mode is dominant to the final stage where the axisymmetry of
the dynamo is almost restored. It is found that magnetic helicity is
transferred forth and back over the spectral space during the
transient phase. Also our simulations shows that the non-axisymmetric distributions of magnetic
helicity tend to follows the regions of the Hale polarity rule.

\end{abstract}

\section{Introduction}

Conservation of the magnetic helicity is significant for many physical
process above and beneath the solar photosphere. After seminal papers
\cite{pouquet-al:1975a} and   \cite{pouquet-al:1975b} it was
understood that the magnetic helicity is one of the key parameters
which determine generation and evolution of the large-scale magnetic
field in the solar dynamo. It is commonly believed that the solar
magnetic fields are generated by the axisymmetric hydromagnetic
dynamo instability in the solar convection zone due to the differential
rotation and helical convective motions. Effects of magnetic helicity
conservation in the axisymmetric dynamo inside convection zone and their impact
on the activity in the regions above the photosphere are lively debated
in the current literature (see, e.g., \cite{brsu05}). Despite the
strong axisymmetry of solar magnetic activity on the long time-scale,
deviations from the axisymmetry are rather strong at any particular
moment of observations. 

For the physical conditions of the modern Sun it was found that the
large-scale non-axisymmetric magnetic field is linearly stable
\cite{rad86AN}. Using the nonlinear mean-field model, in \cite{pk15} was  found that the non-axisymmetric magnetic
perturbations can result to transients in evolution of the axisymmetric
fields if the perturbations are anchored to level of the rotational
subsurface shear layer ($\approx0.9R$). After relaxation of the non-axisymmetric
magnetic perturbation the non-linear dynamo processes can maintain
a weak non-axisymmetric field in expense of the axisymmetric 
magnetic field. It was also found that that the magnetic helicity
conservation affects the interaction of the non-axisymmetric and axisymmetric magnetic fields.

The goal of the paper is to study the evolution of the magnetic helicity
density using the non-axisymmetric mean-field dynamo model. Also, we are interesting to
study redistribution of magnetic helicity density over the partial azimuthal
modes (see \cite{KR80}) of the large-scale magnetic field. The model was
described in details in \cite{pk15}.
In the next two sections we briefly outline the basic equations of our
model and discuss mechanisms of the non-linear interactions between
the axisymmetric andd non-axisymmetric dynamos. The Section 4 contain the description of main results,
discussion and the Section 5 summarizes conclusions of the study.

\section{Basic equations}

Evolution of the large-scale magnetic field in perfectly conductive
media is described by the mean-field induction equation \cite{KR80}:
\begin{equation}
\partial_{t}\left\langle \mathbf{B}\right\rangle =\boldsymbol{\nabla}\times\left(\mathbf{\boldsymbol{\boldsymbol{\mathcal{E}}}+}\left\langle \mathbf{U}\right\rangle \times\left\langle \mathbf{B}\right\rangle \right)\label{eq:mfe}
\end{equation}
where $\boldsymbol{\mathcal{E}}=\left\langle \mathbf{u\times b}\right\rangle $
is the mean electromotive force; $\mathbf{u}$ and $\mathbf{b}$ are
the turbulent fluctuating velocity and magnetic field respectively;
and $\left\langle \mathbf{U}\right\rangle $ and $\left\langle \mathbf{B}\right\rangle $
are the mean velocity and magnetic field. For convenience we decompose
the magnetic field into the axisymmetric, (hereafter $\overline{\mathbf{B}}$-field),
and non-axisymmetric parts, (hereafter $\tilde{\mathbf{B}}$-field):
$\left\langle \mathbf{B}\right\rangle =\overline{\mathbf{B}}+\tilde{\mathbf{B}}$.
We assume that the mean flow is axisymmetric $\left\langle \mathbf{U}\right\rangle \equiv\overline{\mathbf{U}}$.
Let $\hat{\boldsymbol{\phi}}=\mathbf{e_{\phi}}$ and $\hat{\mathbf{r}}=r\mathbf{e}_{r}$
be vectors in the azimuthal and radial directions respectively, then
we represent the mean magnetic field vectors as follows: 
\begin{eqnarray}
\left\langle \mathbf{B}\right\rangle  & = & \overline{\mathbf{B}}+\tilde{\mathbf{B}}\label{eq:b0}\\
\mathbf{\overline{B}} & = & \hat{\boldsymbol{\phi}}B+\nabla\times\left(A\hat{\boldsymbol{\phi}}\right)\label{eq:b1}\\
\tilde{\mathbf{B}} & = & \boldsymbol{\nabla}\times\left(\hat{\mathbf{r}}T\right)+\boldsymbol{\nabla}\times\boldsymbol{\nabla}\times\left(\hat{\mathbf{r}}S\right),\label{eq:b2}
\end{eqnarray}
where $A$, $B$, $T$ and $S$ are scalar functions representing
the axisymmetric and non-axisymmetric parts respectively. Assuming that $A$ and $B$ do not
depend on longitude, Eqs(\ref{eq:b1}, \ref{eq:b2}) ensure that the
field $\left\langle \mathbf{B}\right\rangle $ is divergence-free.
The integration domain includes the solar convection zone from $0.71$
to $0.99R_{\odot}$. The distribution of the mean flows is given by
helioseismology (\cite{Howe2011JPh}). Profiles of the angular
velocity is the same as in \cite{pk15}, (see Figure 1 there).
For the sake of simplicity we neglect the meridional circulation in
the model.

We use formulation for the mean electromotive force obtained in form:
\begin{equation}
\mathcal{E}_{i}=\left(\alpha_{ij}+\gamma_{ij}\right)\left\langle B\right\rangle _{j}-\eta_{ijk}\nabla_{j}\left\langle B\right\rangle _{k}.\label{eq:EMF-1}
\end{equation}
where symmetric tensor $\alpha_{ij}$ models the generation of magnetic
field by the $\alpha$- effect; antisymmetric tensor\textbf{ }$\gamma_{ij}$
controls the mean drift of the large-scale magnetic fields in turbulent
medium; tensor $\eta_{ijk}$ governs the turbulent diffusion. We take
into account the effect of rotation and magnetic field on the mean-electromotive
force (see, e.g., \cite{pi15M} for details). To determine unique
solution we apply the following gauge\textbf{ }(see, e.g., \cite{KR80}):
\begin{equation}
\int_{0}^{2\pi}\int_{-1}^{1}Sd\mu d\phi=0,\,\,\int_{0}^{2\pi}\int_{-1}^{1}Td\mu d\phi=0,\label{eq:norm}
\end{equation}
where $\mu=\cos\theta$ and $\theta$ is the polar angle..

\section{Nonlinear interaction of the axisymmetric and non-axisymmetric modes}

Interaction between the axisymmetric and non-axisymmetric modes in the mean-field dynamo models
can be due to nonlinear dynamo effects, for example, the $\alpha$-effect,
\cite{radler90,moss99}. In our model the $\alpha$ effect takes
into account the kinetic and magnetic helicities in the following
form: 
\begin{eqnarray}
\alpha_{ij} & = & C_{\alpha}\sin^{2}\theta\psi_{\alpha}(\beta)\alpha_{ij}^{(H)}\eta_{T}+\alpha_{ij}^{(M)}\frac{\left\langle \chi\right\rangle \tau_{c}}{4\pi\overline{\rho}\ell^{2}}\label{alp2d}
\end{eqnarray}
where $C_{\alpha}$ is a free parameter which controls the strength
of the $\alpha$- effect due to turbulent kinetic helicity; $\alpha_{ij}^{(H)}$
and $\alpha_{ij}^{(M)}$ express the kinetic and magnetic helicity
parts of the $\alpha$-effect, respectively; $\eta_{T}$ is the magnetic
diffusion coefficient, and $\left\langle \chi\right\rangle
=\left\langle \mathbf{a}\cdot\mathbf{b}\right\rangle $ the helicity
density of the small-scale magnetic field ( $\mathbf{a}$ and $\mathbf{b}$ are the fluctuating parts of magnetic
field vector-potential and magnetic field vector). Both the $\alpha_{ij}^{(H)}$
and $\alpha_{ij}^{(M)}$ depend on the Coriolis number $\Omega^{*}=4\pi{\displaystyle \frac{\tau_{c}}{P_{rot}}}$,
where $P_{rot}$ is the rotational period, $\tau_{c}$ is the convective
turnover time, and $\ell$ is a typical length of the convective flows
(the mixing length). Function $\psi_{\alpha}(\beta)$ controls the
so-called ``algebraic'' quenching of the $\alpha$- effect where
$\beta=\left|\mathbf{\left\langle B\right\rangle }\right|/\sqrt{4\pi\overline{\rho}u'^{2}}$,
$u'$ is the r.m.s. of the convective velocity. 

The magnetic helicity conservation results to the dynamical quenching
of the dynamo. Contribution of the magnetic helicity to the $\alpha$-effect
is expressed by the second term in Eq.(\ref{alp2d}). The magnetic
helicity density of turbulent field, $\left\langle \chi\right\rangle $,
is governed by the conservation law \cite{pip13M}: 
\begin{equation}
\frac{\partial\left\langle \chi\right\rangle ^{(tot)}}{\partial t}=-\frac{\left\langle \chi\right\rangle }{R_{m}\tau_{c}}-2\eta\left\langle \mathbf{B}\right\rangle \cdot\left\langle \mathbf{J}\right\rangle -\boldsymbol{\nabla\cdot}\boldsymbol{\boldsymbol{\mathcal{F}}}^{\chi},\label{eq:helcon-1}
\end{equation}
where $\left\langle \chi\right\rangle ^{(tot)}=\left\langle \chi\right\rangle +\left\langle \mathbf{A}\right\rangle \cdot\left\langle \mathbf{B}\right\rangle $
is the total magnetic helicity density of the mean and turbulent fields,
$\boldsymbol{\boldsymbol{\mathcal{F}}}^{\chi}=-\eta_{\chi}\boldsymbol{\nabla}\left\langle \chi\right\rangle $
is the diffusive flux of the turbulent magnetic helicity, and $R_{m}$
is the magnetic Reynolds number. The coefficient of the turbulent
helicity diffusivity, $\eta_{\chi}$, is chosen ten times smaller
than the isotropic part of the magnetic diffusivity \cite{mitra10}:
$\eta_{\chi}=\frac{1}{10}\eta_{T}$. Similarly to the magnetic field,
the mean magnetic helicity density can be formally decomposed into
the axisymmetric and non-axisymmetric parts: $\left\langle \chi\right\rangle ^{(tot)}=\overline{\chi}^{(tot)}+\tilde{\chi}^{(tot)}$.
The same can be done for the magnetic helicity density of the turbulent
field: $\left\langle \chi\right\rangle =\overline{\chi}+\tilde{\chi}$,
where $\overline{\chi}=\overline{\mathbf{a}\cdot\mathbf{b}}$ and
$\tilde{\chi}=\tilde{\left\langle \mathbf{a}\cdot\mathbf{b}\right\rangle }$.
Then we have, 
\begin{align}
\overline{\chi}^{(tot)} & =\overline{\chi}+\overline{\mathbf{A}}\cdot\overline{\mathbf{B}}+\overline{\tilde{\mathbf{A}}\cdot\tilde{\mathbf{B}}},\label{eq:t1}\\
\tilde{\chi}^{(tot)} & =\tilde{\chi}+\overline{\mathbf{A}}\cdot\tilde{\mathbf{B}}+\tilde{\mathbf{A}}\cdot\overline{\mathbf{B}}+\tilde{\mathbf{A}}\cdot\tilde{\mathbf{B}},\label{eq:t2}
\end{align}
Evolution of the $\overline{\chi}$ and $\tilde{\chi}$ is governed
by the corresponding parts of Eq(\ref{eq:helcon-1}). Thus, the model
takes into account contributions of the axisymmetric and non-axisymmetric fields in the whole
magnetic helicity density balance, providing a non-linear coupling.
We see that the $\alpha$-effect is dynamically linked to the longitudinally
averaged magnetic helicity of the non-axisymmetric $\tilde{\mathbf{B}}$-field,
which is the last term in Eq(\ref{eq:t1}). Thus, the nonlinear $\alpha$-effect
is non-axisymmetric, and it results in coupling between the axisymmetric and
non-axisymmetric modes. The coupling works in both directions. For instance, the
azimuthal $\alpha$-effect results in $\mathcal{E}_{\phi}=\alpha_{\phi\phi}\left\langle B_{\phi}\right\rangle $.
If we denote the non-axisymmetric part of the $\alpha_{\phi\phi}$ by $\tilde{\alpha}_{\phi\phi}$
then the mean electromotive force is $\overline{\mathcal{E}}_{\phi}=\overline{\alpha}_{\phi\phi}\overline{B}_{\phi}+\overline{\tilde{\alpha}_{\phi\phi}\tilde{B}_{\phi}}$.
This introduces a new generation source which is usually ignored in
the axisymmetric dynamo models. 

The numerical scheme employs the spherical harmonics decomposition
for the non-axisymmetric part of the problem and the finite differences
in radial direction. The axisymmetric part of the problem was integrated using
the pseudo-spectral method in latitudinal direction and the finite
differences in radial direction. The parameters of the solar convection
zone are taken from Stix\cite{stix:02} and they were described in \cite{pk15}. At
the bottom of the convection zone we set up a perfectly conducting
boundary condition for the axisymmetric magnetic field, and for the
non-axisymmetric field we set the functions $S$ and $T$ to zero.
At the top of the convection zone the poloidal field is smoothly matched
to the external potential field. The magnetic helicity conservation
is determined by the magnetic Reynolds number $R_{m}$. In this paper
we employ $R_{m}=10^{4}$.

The axisymmetric field was started from a developed non-linear stage. This stage
is characterized by the established oscillating dynamo waves drifting
from the bottom to the top of the convection zone. In subsurface shear
layer the wave of the axisymmetric toroidal magnetic field is deflected equator-ward.
The model satisfactory reproduce the sunspot activity time-latitude
diagram and reversals of the polar magnetic fields as results of dynamo
wave of the axisymmetric poloidal magnetic fields.

Previously, it was found that in the nonlinear non-axisymmetric solar dynamo the
non-axisymmetric magnetic field relax to the stage where its energy is about factor
$10^{-6}$ off the equipartition level of the energy of the convective
motions. However, the strong non-axisymmetric magnetic field could be maintained
by the periodic seed from a decaying active regions or other process.
In the paper we do not concern this important issue. Instead we restrict
our study by the case of the single perturbation. For the seed field
we consider a non-symmetric relative to the equator perturbation represented
by a sum of the equatorial dipole (l=1, m=$\pm$1,2) and quadrupole
($l=2$, $m=\pm1,2$) components (see \cite{pk15}). We employ the initial
non-axisymmetric perturbation to be concentrated in the near-surface shear layer. 

\begin{figure}
\includegraphics[width=1\textwidth]{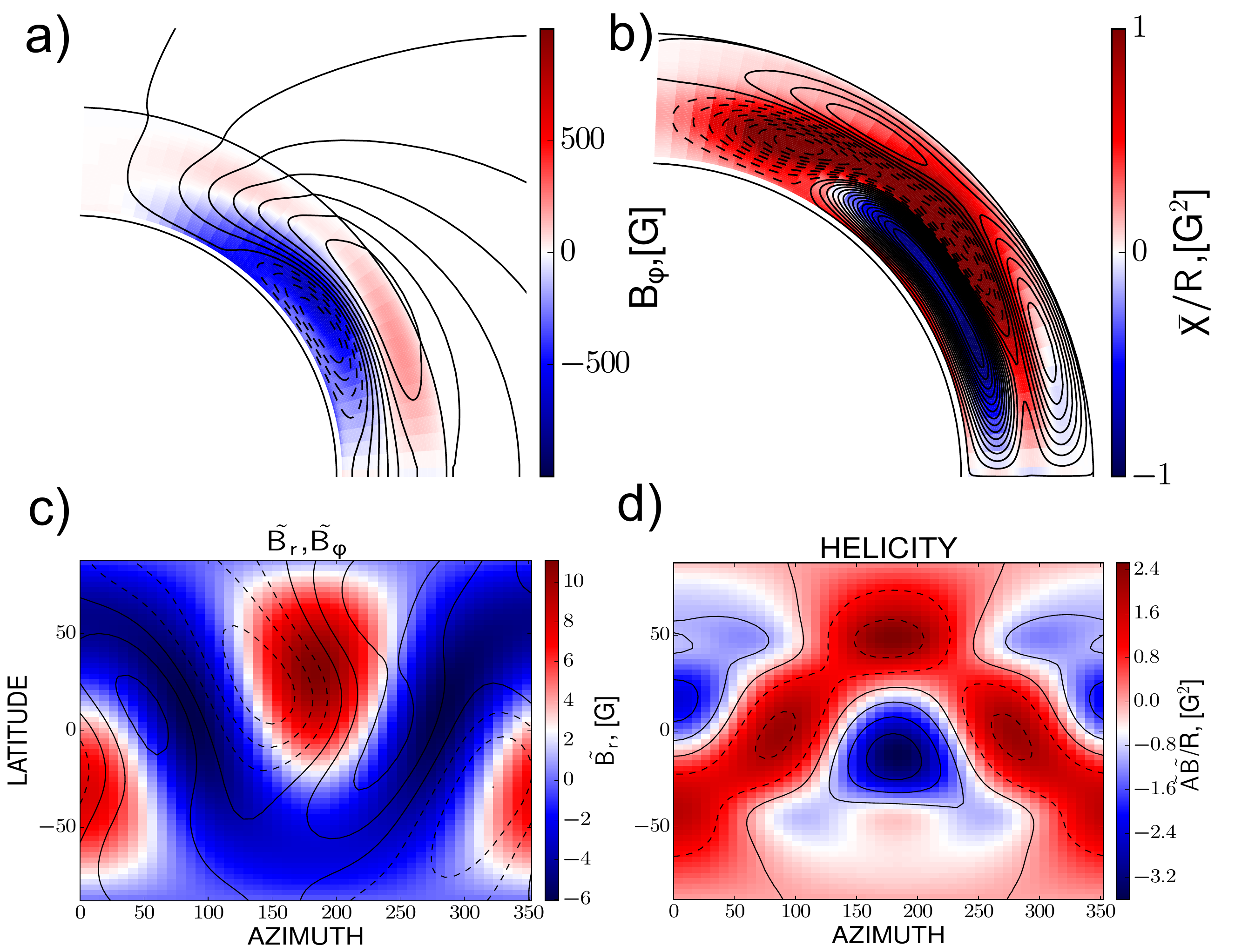}

\caption{\label{fig:init}. Distribution of the axisymmetric and non-axisymmetric magnetic field just
after initialization : a) the axisymmetric magnetic field in meridional cross-section,
the geometry of the poloidal field is shown by contours; b) Color
image shows the magnetic helicity density $\bar{\chi}$ and the \textbf{$\overline{\mathbf{A}}\cdot\overline{\mathbf{B}}$}
is shown by contours. It varies in the same range as the $\bar{\chi}$
; c) color image shows the non-axisymmetric $\tilde{B_{r}}$ at the surface and
contours show the azimuthal component of the non-axisymmetric magnetic field, $\tilde{B}_{\phi}$;
d) color image shows distribution of the $\tilde{\mathbf{A}}\cdot\tilde{\mathbf{B}}$
and contours are for the non-axisymmetric part of the small-scale magnetic helicity
density, $\tilde{\chi}$, which varies in the same interval of magnitude
as the $\tilde{\mathbf{A}}\cdot\tilde{\mathbf{B}}$ .}
\end{figure}

\section{Results and discussion}

Distribution of the axisymmetric and non-axisymmetric magnetic field just after initialization
is shown by Figure1(a,c). The dynamic of the system is started to
evolve at the epoch of the solar minimum when the polar field has
maximum amplitude and the strength of the toroidal field in the near-surface
layer is minimal. Figure 1(b,c) show distributions of the magnetic
helicity density for the axisymmetric and the non-axisymmetric parts respectively. At the
Northern hemisphere the axisymmetric part of the small-scale magnetic helicity
is negative near the bottom of the convection where the dynamo wave
of the axisymmetric toroidal magnetic field zone is concentrated. Also we have
the negative patch of the $\bar{\chi}$ at the near surface where
the dynamo wave from the previous cycle is decaying. Note that the
$\overline{\mathbf{A}}\cdot\overline{\mathbf{B}}$ is positive at
the North and the sign of it is in balance with the sign of the $\bar{\chi}$
because of the helicity conservation in the model. 
At the initialization time the non-axisymmetric perturbation is of the strength
about 10G at the surface and it decays to zero at the $r=0.9R$. Form
Figure 1 it is seen that the injected helicity density of the non-axisymmetric magnetic
field is about factor 2 larger than the helicity density of the axisymmetric
part of the magnetic field.. 

\begin{figure}
\includegraphics[width=1\textwidth]{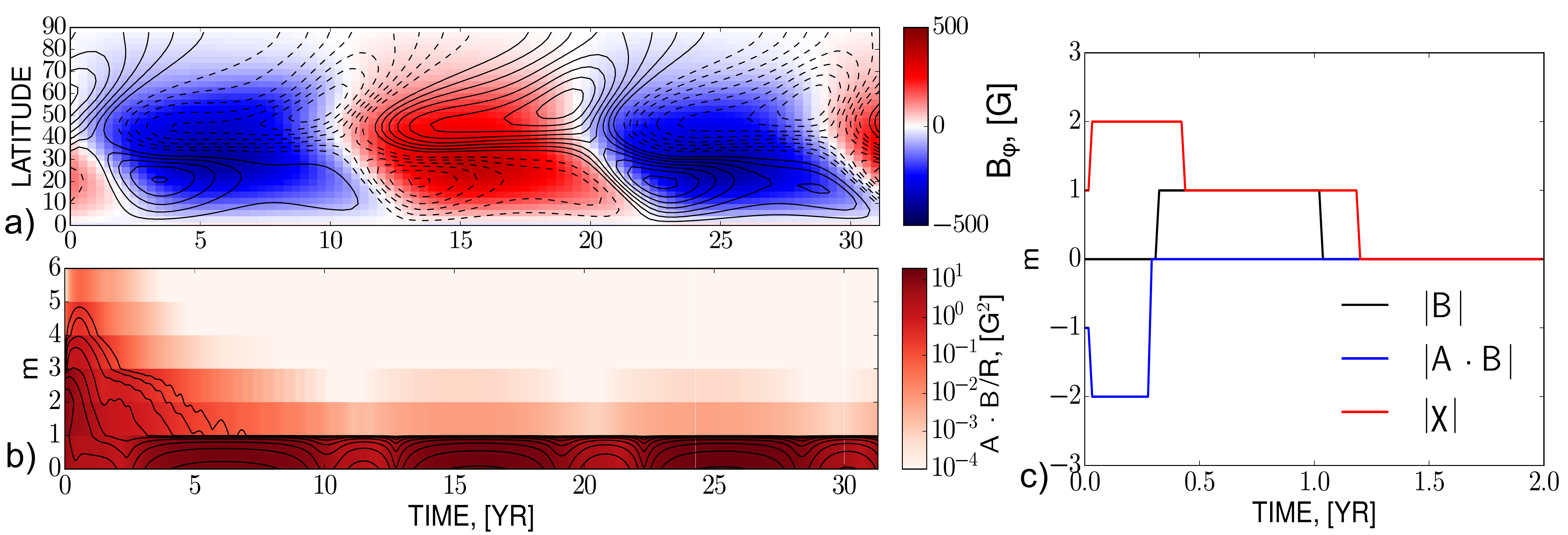}

\caption{\label{fig:Time-latitude-diagram-(a),}. a) Time-latitude diagram
of the axisymmetric toroidal magnetic field at the r=0.9R (color image), contours
show the radial magnetic field at the surface (they vary in range
$\pm5$G); b) Color image show variations of the integral of the $\left|\mathbf{\left\langle A\right\rangle \cdot}\left\langle \mathbf{B}\right\rangle \right|$
over latitudes with time at the r=0.9R. The first six partial dynamo
modes are shown (m=0 corresponds to the axisymmetric magnetic field). Contours
show the same for the small-scale helicity density, which varies within
the same range of magnitudes as the $\left|\mathbf{\left\langle A\right\rangle \cdot}\left\langle \mathbf{B}\right\rangle \right|$;
c) The number of the partial non-axisymmetric mode which has the strongest magnitude
of the magnetic field (black line), the helicity density of the large-scale
magnetic field (blue line, the number was reverted to avoid the overlap),
and the small-scale magnetic helicity density (red line).}
\end{figure}

Figure 2(a) shows the time-latitude evolution of the axisymmetric magnetic field
near the solar surface. Note that in compare with results of the previous
paper, \cite{pk15}, the butterfly diagram is unchanged after initialization
of the non-axisymmetric perturbation because the initialization time in these simulations
corresponds to epoch of the minimum of the dynamo cycle. Figure 2(b)
shows variations of the $\int\left|\left\langle \mathbf{A}\right\rangle \cdot\left\langle \mathbf{B}\right\rangle \right|d\mu$
at the radial distance $r=0.95R$. The first six partial dynamo modes
are shown (m=0 corresponds to the axisymmetric magnetic field). It is seen that
after initialization the helicity is transferred from the non-axisymmetric modes
to the axisymmetric magnetic field. Figure 2(c) gives a more detailed information
about redistribution of the magnetic energy and magnetic helicity
among the partial modes of the non-axisymmetric magnetic field in the dynamo for
period of the first two years after initialization. It is seen that
for the short period about half an year after initialization the non-axisymmetric
m=1 dynamo mode become the dominant. This happens after the maximum
of the helicity density transfer to the m=2 modes.

The strength of the non-axisymmetric magnetic field decays at the time about 5 year
to the level which is about $10^{-3}$off the strength of the axisymmetric magnetic
field. The nonlinear interaction of the axisymmetric and non-axisymmetric magnetic fields
via magnetic helicity helps to maintain the non-axisymmetric magnetic field from
a complete decay. 

Figure 3,4 show the snapshot of magnetic helicity density and the
large-scale magnetic field distributions at the surface and configurations
of the magnetic field above the photosphere during the transient phase
(at the time of one year after initialization) and three years after.
The initialization epoch corresponds to the minimum of the dynamo
cycle. Due to this fact the dynamo model demonstrate inversions of
the hemispheric helicity rule. This is supported from observations
\cite{zetal10,2013ApJ...772...52G} and axisymmetric dynamo models
\cite{pip13M,sok2013}. It is interesting that the non-axisymmetric
helicity density distribution tends to follow to patterns of the large-scale
non-axisymmetric magnetic field during evolution. Therefore, according to our model
the change of the helicity sign in the longitudinal direction could
occurs at the sectoral boundaries of the large-scale magnetic field.
During the transient phase of evolution the axisymmetric toroidal field at the
near surface layer is positive in the Northern hemisphere and it is
negative in the Southern hemisphere. The Figure 3(b) shows that
in the Northern hemisphere the helicity density changes the sign from
negative to positive at the anti-Hale sectoral boundary (\cite{sval11})
and in opposite direction at the sectoral boundary which correspond
to the Hale polarity rule for the given cycle.  This
is probably because of the local balance of the magnetic helicity for the large and
small scales which is prescribed by simple anzatz given by Eq.(8). It
is interesting to note that the flare activity is concentrated near
the anti-Hale sectorial boundaries \cite{sval11}.

Figure 4 shows that
about three years after initialization the model restores the dominance
of the axisymmetric magnetic fields. The hemispheric helicity sign rule is also
restored to the normal case when the Northern hemisphere has the negative
sign of the magnetic helicity and the Southern has positive one. 

\begin{figure}
\includegraphics[width=1.\textwidth]{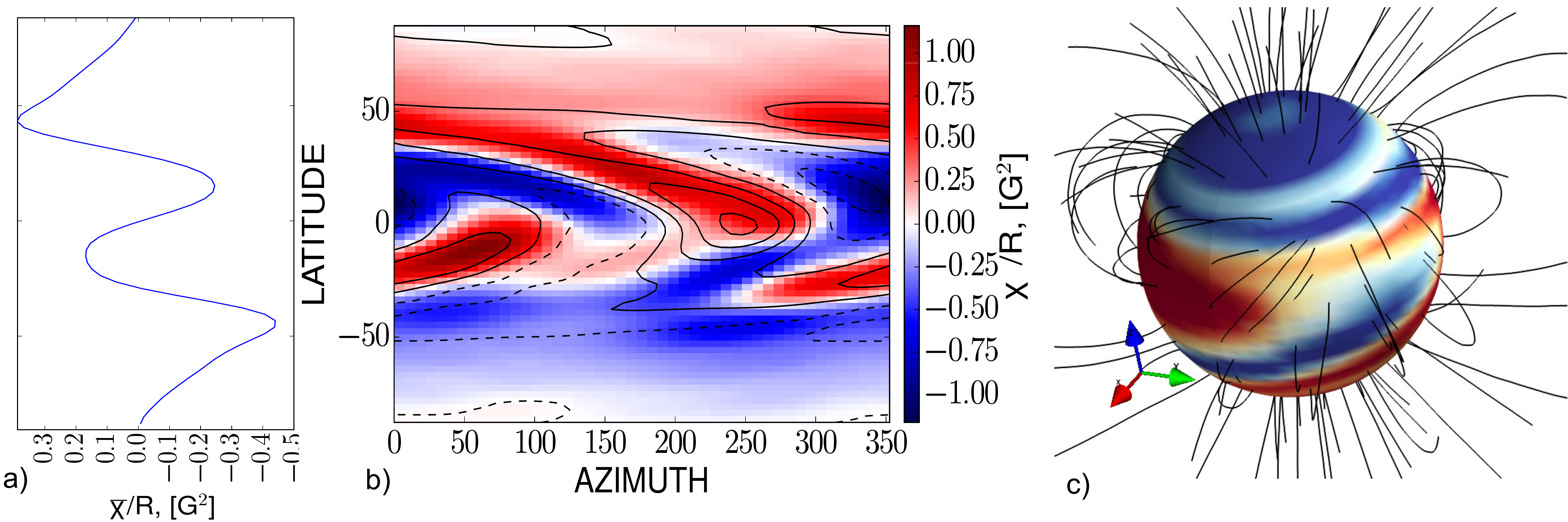}\caption{Snapshots of the magnetic helicity density and magnetic field distributions
at the time one year after initialization of the non-axisymmetric perturbation:
a) mean small-scale helicity density, $\bar{\chi}$; b) color image
shows the sum of the non-axisymmetric and axisymmetric parts of the magnetic helicity density,
$\left\langle \chi\right\rangle =\overline{\chi}+\tilde{\chi}$, contours
show the strength of the radial magnetic field (within the range of
$\pm10$G);c) shows magnetic field lines and color image shows the
strength of the radial magnetic field. }
\end{figure}

\begin{figure}
\includegraphics[width=1.\textwidth]{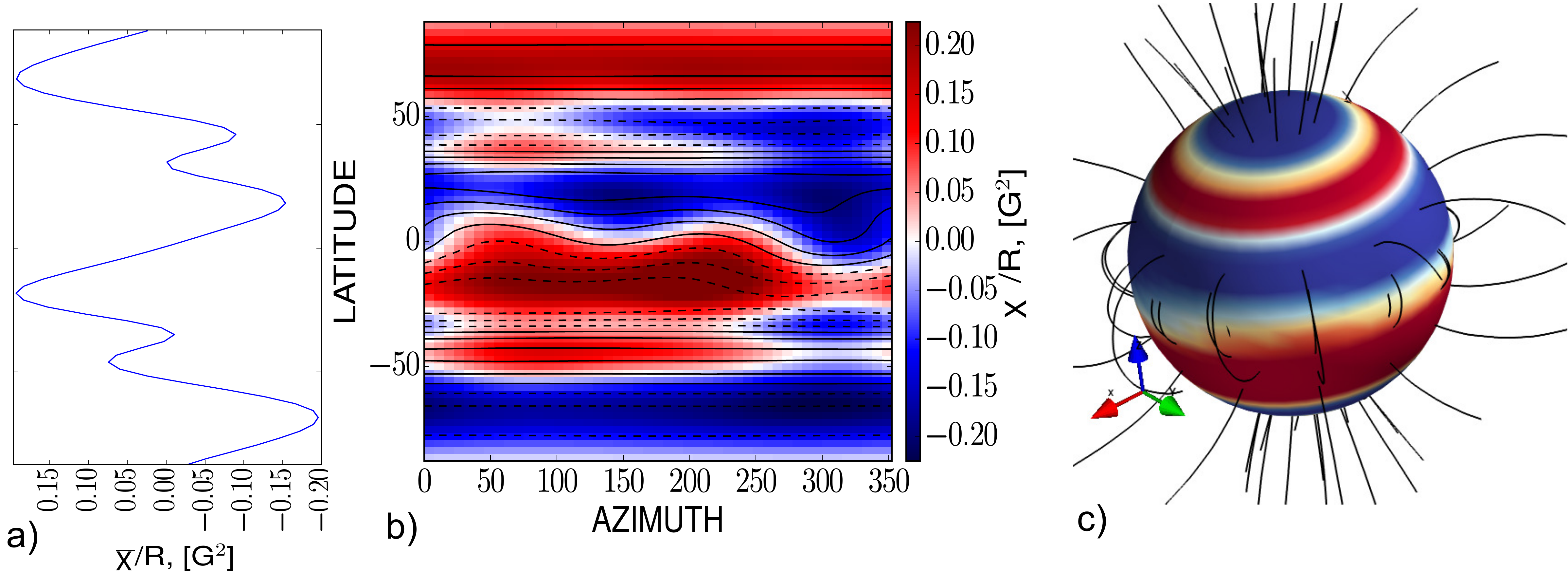}\caption{The same as Figure 3 for the time three years after initialization
of the non-axisymmetric perturbation. }
\end{figure}

\section{Conclusions }

The main goal of the paper was to illustrate some effects of the magnetic
helicity conservation in the non-axisymmetric dynamo model. It was
found that the non-axisymmetric perturbations goes through the transient phase when
the maximum of the magnetic energy in the near surface layer shift
from the m=0 dynamo mode (the axisymmetric magnetic field) to the m=1 mode.
At the same time the magnetic helicity density is transferred over
the spectral modes in the same direction at the beginning of the transient
phase and back shortly after while. Our simulations shows that the
non-axisymmetric distributions of the small-scale  magnetic
helicity tend to follows the regions of the Hale polarity rule. The study is restricted
to the case of the single perturbation at the particular moment of
the dynamo cycle. Thus it would be interesting to extend our investigation
for a more general case including a more realistic form of the non-axisymmetric perturbations. 

Acknowledgments The work supported by RFBR under grants 14-02-90424,
15-02-01407 and the project II.16.3.1 of ISTP SB RAS.


\end{document}